\documentclass[11pt]{article}
\usepackage{amsmath,amssymb,amsthm,amsxtra,overpic,bbm,bm,epsfig,subfigure}
\usepackage{color}
\usepackage{multirow}

\def\thefootnote{\fnsymbol{footnote}}

\begin{document}

\vspace{0.2cm}

\begin{center}
{\Large\bf Resolving the octant of $\theta^{}_{23}$ via radiative
$\mu$-$\tau$ symmetry breaking}
\end{center}

\vspace{0.2cm}

\begin{center}
{\bf Shu Luo $^{a}$} \footnote{E-mail: luoshu@xmu.edu.cn}
\quad
{\bf Zhi-zhong Xing $^{b,c}$} \footnote{E-mail: xingzz@ihep.ac.cn}
\\
{$^a$Department of Astronomy and Institute of Theoretical Physics
and Astrophysics,\\ Xiamen University, Xiamen, Fujian 361005, China \\
$^b$Institute of High Energy Physics, Chinese Academy of
Sciences, Beijing 100049, China \\
$^c$Center for High Energy Physics, Peking University, Beijing 100080, China}
\end{center}

\vspace{1.5cm}
\begin{abstract}
We point out that the observed neutrino mixing pattern at low
energies is very likely to originate from the $3\times 3$
Pontecorvo-Maki-Nakagawa-Sakata (PMNS) lepton flavor mixing matrix
$U$ which possesses the exact $\mu$-$\tau$ permutation symmetry
$|U^{}_{\mu i}| = |U^{}_{\tau i}|$ (for $i=1,2,3$) at a superhigh
energy scale $\Lambda^{}_{\mu\tau} \sim 10^{14}$ GeV. The deviation
of $\theta^{}_{23}$ from $45^\circ$ and that of $\delta$ from
$270^\circ$ in the standard parametrization of $U$ are therefore a
natural consequence of small PMNS $\mu$-$\tau$ symmetry breaking via
the renormalization-group equations (RGEs) running from
$\Lambda^{}_{\mu\tau}$ down to the electroweak scale
$\Lambda^{}_{\rm EW} \sim 10^2$ GeV. In fitting current experimental
data we find that the RGE-corrected value of $\theta^{}_{23}$ is
uniquely correlated with the neutrino mass ordering: $\theta^{}_{23}
\simeq 42.4^\circ$ reported by Capozzi {\it et al} (or
$\theta^{}_{23} \simeq 48.9^\circ$ reported by Forero {\it et al})
at $\Lambda^{}_{\rm EW}$ can arise from $\theta^{}_{23} = 45^\circ$
at $\Lambda^{}_{\mu\tau}$ in the minimal supersymmetric standard
model if the neutrino mass ordering is inverted (or normal).
Accordingly, the preliminary best-fit results of $\delta$ at
$\Lambda^{}_{\rm EW}$ can also evolve from $\delta = 270^\circ$ at
$\Lambda^{}_{\mu\tau}$ no matter whether the massive neutrinos are
Dirac or Majorana particles.
\end{abstract}

\begin{flushleft}
\hspace{0.8cm} PACS number(s): 14.60.Pq, 13.10.+q, 25.30.Pt
\end{flushleft}

\def\thefootnote{\arabic{footnote}}
\setcounter{footnote}{0}

\newpage

\section{Introduction}

From the discovery of atmospheric neutrino oscillations in 1998
\cite{SK} until the observation of the smallest neutrino mixing
angle $\theta^{}_{13}$ in 2012 \cite{DYB}, experimental neutrino
physics was in full flourish. Today the era of precision
measurements has come. A number of undergoing and upcoming neutrino
oscillation experiments aim to determine the neutrino mass ordering,
to probe the octant of the largest neutrino mixing angle
$\theta^{}_{23}$, and to measure the Dirac CP-violating phase
$\delta$. Such knowledge will be fundamentally important, as it can
help identify the underlying flavor symmetry or dynamics behind the
observed pattern of lepton flavor mixing.

As far as the octant of $\theta^{}_{23}$ is concerned, it is
desirable to know whether this ``atmospheric neutrino mixing" angle
deviates from $45^\circ$ or not, and if it does, how large or small the
deviation is and in what direction the deviation evolves. A global
analysis of current neutrino oscillation data done by Capozzi {\it
et al} \cite{Fogli} yields the best-fit result $\theta^{}_{23}
\simeq 41.4^\circ$ (normal neutrino mass ordering) or
$\theta^{}_{23} \simeq 42.4^\circ$ (inverted neutrino mass
ordering), which has a preference for the first octant (i.e.,
$\theta^{}_{23} < 45^\circ$). In contrast, another best-fit result
reported by Forero {\it et al} \cite{Valle} is $\theta^{}_{23}
\simeq 48.8^\circ$ (normal ordering) or $\theta^{}_{23} \simeq
49.2^\circ$ (inverted ordering), by which the second octant (i.e.,
$\theta^{}_{23} > 45^\circ$) is favored. In both cases
$\theta^{}_{23} = 45^\circ$ will be allowed when the $1\sigma$ or
$2\sigma$ error bars are taken into account. Hence the octant of
$\theta^{}_{23}$ remains an open issue, and a resolution to this
puzzle awaits more accurate experimental data.

On the other hand, the best-fit results of $\delta$ in both Ref.
\cite{Fogli} and Ref. \cite{Valle} are close to an especially
interesting value, $270^\circ$, although the confidence level
remains quite low. In fact, the T2K measurement of a relatively
strong $\nu^{}_\mu \to \nu^{}_e$ appearance signal \cite{T2K} plays
a crucial role in the global fit to make $\theta^{}_{13}$ consistent
with the Daya Bay result \cite{DYB} and drive a slight but
intriguing preference for $\delta \simeq 270^\circ$
\cite{Fogli,Valle}. If this preliminary expectation turns out to be
true, there will be no problem to observe significant effects of
leptonic CP violation in the forthcoming long-baseline neutrino
oscillation experiments.

On the theoretical side, $\theta^{}_{23} = 45^\circ$ and $\delta =
270^\circ$ are a straightforward consequence of the $\mu$-$\tau$
permutation symmetry manifesting itself in the $3\times 3$
Pontecorvo-Maki-Nakagawa-Sakata (PMNS) lepton flavor mixing matrix
$U$ \cite{MNS}: $|U^{}_{\mu i}| = |U^{}_{\tau i}|$ (for $i=1,2,3$),
which can easily be embedded in an explicit flavor symmetry model.
Hence the deviation of $\theta^{}_{23}$ from $45^\circ$ and that of
$\delta$ from $270^\circ$ must be related to small PMNS $\mu$-$\tau$
symmetry breaking effects. This observation is important and
suggestive, implying that the observed pattern of the PMNS matrix
$U$ should have an approximate $\mu$-$\tau$ symmetry of the form
$|U^{}_{\mu i}| \simeq |U^{}_{\tau i}|$ at low energies \cite{XZ}.
In comparison, the Cabibbo-Kobayashi-Maskawa (CKM) quark flavor
mixing matrix $V$ \cite{CKM} does not possess such a peculiar
structure.

In this work we pay particular attention to a very real possibility:
the PMNS $\mu$-$\tau$ symmetry is exact at a superhigh energy scale
$\Lambda^{}_{\mu\tau}$ where both tiny neutrino masses and large
neutrino mixing angles could naturally be explained in a
well-founded theoretical framework (e.g., with the canonical seesaw
mechanism \cite{SS} and proper flavor symmetry groups
\cite{Flavor}). In this case we find that it is possible to resolve
the octant of $\theta^{}_{23}$ and the quadrant of $\delta$ via
radiative $\mu$-$\tau$ symmetry breaking effects. Namely, the
equalities $|U^{}_{\mu i}| = |U^{}_{\tau i}|$ are more or less
violated when they evolve from $\Lambda^{}_{\mu\tau} \sim 10^{14}$
GeV down to the electroweak scale $\Lambda^{}_{\rm EW} \sim 10^2$
GeV via the relevant renormalization-group equations (RGEs), such
that the correct octant of $\theta^{}_{23}$ and the correct quadrant
of $\delta$ can consequently be obtained. We carry out a numerical
analysis of the issue for both Dirac and Majorana neutrinos based on
the one-loop RGEs in the minimal supersymmetric standard model
(MSSM)
\footnote{In this connection the standard-model RGEs are less
interesting for two reasons: (a) it will be difficult to make the
deviation of $\theta^{}_{23}$ from $45^\circ$ appreciable even if
the neutrino masses are nearly degenerate; (b) the standard model itself
will largely suffer from the vacuum-stability problem for the measured value
of the Higgs mass ($\simeq 125$ GeV) as the energy scale is above
$10^{10}$ GeV \cite{XZZ}.}.
A striking finding of ours in fitting current neutrino oscillation
data is that the RGE-corrected value of $\theta^{}_{23}$ is
uniquely correlated with the neutrino mass ordering:
$\theta^{}_{23} \simeq 42.4^\circ$ reported in Ref. \cite{Fogli} (or
$\theta^{}_{23} \simeq 48.9^\circ$ reported in Ref. \cite{Valle})
at $\Lambda^{}_{\rm EW}$ can evolve from $\theta^{}_{23} = 45^\circ$
at $\Lambda^{}_{\mu\tau}$ only when the neutrino masses have an
inverted (or normal) ordering. Accordingly, the preliminary best-fit
results of $\delta$ at $\Lambda^{}_{\rm EW}$ can also originate from
$\delta = 270^\circ$ at $\Lambda^{}_{\mu\tau}$ thanks to radiative
$\mu$-$\tau$ symmetry breaking. Such remarkable
results are independent of any specific models of neutrino mass
generation and lepton flavor mixing, and they will soon be tested in
the upcoming precision experiments of neutrino oscillations.

\section{The RGEs of $\mu$-$\tau$ symmetry breaking}

The PMNS lepton flavor mixing matrix can be parametrized in the
following ``standard" way \cite{PDG}:
\begin{eqnarray}
U = \left( \begin{matrix} U^{}_{e 1} & U^{}_{e 2} & U^{}_{e
3} \cr U^{}_{\mu 1} & U^{}_{\mu 2} & U^{}_{\mu 3} \cr U^{}_{\tau 1}
& U^{}_{\tau 2} & U^{}_{\tau 3} \cr \end{matrix} \right)
= \left( \begin{matrix} c^{}_{12} c^{}_{13} & s^{}_{12} c^{}_{13}
& s^{}_{13} e^{-{\rm i} \delta} \cr -s^{}_{12} c^{}_{23} - c^{}_{12}
s^{}_{13} s^{}_{23} e^{{\rm i} \delta} & c^{}_{12} c^{}_{23} -
s^{}_{12} s^{}_{13} s^{}_{23} e^{{\rm i} \delta} & c^{}_{13}
s^{}_{23} \cr s^{}_{12} s^{}_{23} - c^{}_{12} s^{}_{13} c^{}_{23}
e^{{\rm i} \delta} & -c^{}_{12} s^{}_{23} - s^{}_{12} s^{}_{13}
c^{}_{23} e^{{\rm i} \delta} & c^{}_{13} c^{}_{23} \cr
\end{matrix} \right) P^{}_\nu \; ,
\end{eqnarray}
where $c^{}_{ij} \equiv \cos\theta^{}_{ij}$, $s^{}_{ij} \equiv
\sin\theta^{}_{ij}$ (for $ij = 12, 13, 23$), $\delta$ is referred to
as the Dirac CP-violating phase, and $P^{}_\nu = {\rm
Diag}\left\{e^{{\rm i}\rho}, e^{{\rm i}\sigma}, 1\right\}$ contains
two extra phase parameters if massive neutrinos are the Majorana
particles. Up to now $\theta^{}_{12}$, $\theta^{}_{13}$ and
$\theta^{}_{23}$ have all been measured to a good degree of
accuracy, and some preliminary hints for a nontrivial value of
$\delta$ have also been obtained from a global analysis of current
neutrino oscillation data \cite{Fogli,Valle}. Here we are concerned
about the three PMNS $\mu$-$\tau$ ``asymmetries":
\begin{eqnarray}
\Delta^{}_1 \hspace{-0.2cm} & \equiv & \hspace{-0.2cm} |U^{}_{\tau
1}|^2 - |U^{}_{\mu 1}|^2 = \left( \cos^2\theta^{}_{12}
\sin^2\theta^{}_{13} - \sin^2\theta^{}_{12} \right) \cos
2\theta^{}_{23} - \sin 2\theta^{}_{12} \sin\theta^{}_{13} \sin
2\theta^{}_{23} \cos\delta \; , \hspace{0.6cm}
\nonumber \\
\Delta^{}_2 \hspace{-0.2cm} & \equiv & \hspace{-0.2cm} |U^{}_{\tau
2}|^2 - |U^{}_{\mu 2}|^2 = \left( \sin^2\theta^{}_{12}
\sin^2\theta^{}_{13} - \cos^2\theta^{}_{12} \right) \cos
2\theta^{}_{23} + \sin 2\theta^{}_{12} \sin\theta^{}_{13} \sin
2\theta^{}_{23} \cos\delta \; ,
\nonumber \\
\Delta^{}_3 \hspace{-0.2cm} & \equiv & \hspace{-0.2cm} |U^{}_{\tau
3}|^2 - |U^{}_{\mu 3}|^2 = \cos^2\theta^{}_{13} \cos 2\theta^{}_{23}
\; ,
\end{eqnarray}
which satisfy the sum rule $\Delta^{}_1 + \Delta^{}_2 + \Delta^{}_3
=0$. All the three $\Delta^{}_i$ vanish when the exact $\mu$-$\tau$
permutation symmetry holds.

We conjecture that the exact PMNS $\mu$-$\tau$ symmetry (i.e.,
$\Delta^{}_i = 0$) can be realized at $\Lambda^{}_{\mu\tau} \sim
10^{14}$ GeV in a given neutrino mass model with a proper flavor
symmetry group \cite{Flavor}. In view of the facts that a nonzero
and relatively large $\theta^{}_{13} $ has been observed and the
preliminary best-fit value of $\delta$ is not far from $270^\circ$
at the electroweak scale \cite{Fogli,Valle}, we infer that the
condition for all the three $\Delta^{}_i$ to vanish should naturally
be $\theta^{}_{23} = 45^\circ$ and $\delta = 270^\circ$ at the
$\mu$-$\tau$ symmetry scale
\footnote{Although $\Delta^{}_i =0$ might also result from
$\theta^{}_{23} = 45^\circ$ and $\theta^{}_{13} = 0^\circ$ or
$\theta^{}_{23} = 45^\circ$ and $\delta = 90^\circ$ \cite{XZ2008},
neither of them is close to the best-fit results of the
lepton flavor mixing parameters reported in
Refs. \cite{Fogli} and \cite{Valle}. These two possibilities
are much less likely because they have to invoke violent RGE running
effects between $\Lambda^{}_{\mu\tau}$ and $\Lambda^{}_{\rm EW}$ in
order to fit the present experimental data, which actually favor
slight $\mu$-$\tau$ symmetry breaking \cite{XZ}. That is why we concentrate
our interest only on the possibility of $\theta^{}_{23} = 45^\circ$
and $\delta = 270^\circ$ in this paper.}.
In this case $\Delta^{}_i \neq 0$ can
therefore be achieved at $\Lambda^{}_{\rm EW} \sim 10^2$ GeV through
the RGE running effects. The one-loop RGEs of $|U^{}_{\alpha
i}|^2$ (for $\alpha =e, \mu, \tau$ and $i=1,2,3$) have been derived
by one of us in Ref. \cite{Luo}. So it is straightforward to write
out the RGEs of $\Delta^{}_i$ in the MSSM as follows.

\subsection{Dirac neutrinos}

If massive neutrinos are the Dirac particles, we find
\begin{eqnarray}
16\pi^2 \frac{{\rm d}\Delta^{}_1}{{\rm d}t} \hspace{-0.2cm} & = &
\hspace{-0.2cm} -y^2_\tau \left[ \xi^{}_{21} \left( |U^{}_{\tau
1}|^2 \Delta^{}_2 + |U^{}_{\tau 2}|^2 \Delta^{}_1 + |U^{}_{e 3}|^2
\right) + \xi^{}_{31} \left( |U^{}_{\tau 1}|^2 \Delta^{}_3 +
|U^{}_{\tau 3}|^2 \Delta^{}_1 + |U^{}_{e 2}|^2 \right) \right] \; ,
\nonumber \\
16\pi^2 \frac{{\rm d}\Delta^{}_2}{{\rm d}t} \hspace{-0.2cm} & = &
\hspace{-0.2cm} +y^2_\tau \left[ \xi^{}_{21} \left( |U^{}_{\tau
1}|^2 \Delta^{}_2 + |U^{}_{\tau 2}|^2 \Delta^{}_1 + |U^{}_{e 3}|^2
\right) - \xi^{}_{32} \left( |U^{}_{\tau 2}|^2 \Delta^{}_3 +
|U^{}_{\tau 3}|^2 \Delta^{}_2 + |U^{}_{e 1}|^2 \right) \right] \; ,
\nonumber \\
16\pi^2 \frac{{\rm d}\Delta^{}_3}{{\rm d}t} \hspace{-0.2cm} & = &
\hspace{-0.2cm} +y^2_\tau \left[ \xi^{}_{31} \left( |U^{}_{\tau
1}|^2 \Delta^{}_3 + |U^{}_{\tau 3}|^2 \Delta^{}_1 + |U^{}_{e 2}|^2
\right) + \xi^{}_{32} \left( |U^{}_{\tau 2}|^2 \Delta^{}_3 +
|U^{}_{\tau 3}|^2 \Delta^{}_2 + |U^{}_{e 1}|^2 \right) \right] \; ,
\hspace{0.5cm}
\end{eqnarray}
where $t\equiv \ln (\mu/\Lambda^{}_{\mu\tau})$ with $\mu$ being an
arbitrary scale between $\Lambda^{}_{\rm EW}$ and
$\Lambda^{}_{\mu\tau}$, $y^2_\tau = \left(1+ \tan^2\beta\right)
m^2_\tau/v^2$ is the Yukawa coupling eigenvalue of the tau lepton in
the MSSM with $\tan\beta$ and $v$ being self-explaining, and
$\xi^{}_{ij} \equiv (m^2_i + m^2_j)/\Delta m^2_{ij}$ with $\Delta
m^2_{ij} \equiv m^2_i - m^2_j$ being the neutrino mass-squared
differences. Given the fact $|\Delta m^2_{31}| \simeq |\Delta
m^2_{32}| \sim 30 \Delta m^2_{21}$ with $\Delta m^2_{21} \simeq 7.5
\times 10^{-5} ~{\rm eV}^2$, $\xi^{}_{21} \gg |\xi^{}_{31}| \simeq
|\xi^{}_{32}|$ is expected to hold in most cases. But this does not
necessarily mean that the $\mu$-$\tau$ asymmetry $\Delta^{}_3$
should be more stable against radiative corrections than the other
two asymmetries. The reason is simply that the running behaviors of
$\Delta^{}_i$ depend also on the initial inputs of all the nine
$|U^{}_{\alpha i}|^2$. In general, however, an appreciable deviation
of $\theta^{}_{23}$ from $45^\circ$ (i.e., an appreciable deviation
of $\Delta^{}_3$ from zero) requires a sufficiently large value of
$\tan\beta$, and its evolving direction is governed by the neutrino
mass ordering or equivalently the sign of $\Delta m^2_{31}$ or
$\Delta m^2_{32}$.

\subsection{Majorana neutrinos}

If massive neutrinos are the Majorana particles, we arrive at
\begin{eqnarray}
16\pi^2 \frac{{\rm d}\Delta^{}_1}{{\rm d}t} \hspace{-0.2cm} & = &
\hspace{-0.2cm} -y^2_\tau \left\{ \xi^{}_{21} \left( |U^{}_{\tau
1}|^2 \Delta^{}_2 + |U^{}_{\tau 2}|^2 \Delta^{}_1 + |U^{}_{e 3}|^2
\right) + \xi^{}_{31} \left( |U^{}_{\tau 1}|^2 \Delta^{}_3 +
|U^{}_{\tau 3}|^2 \Delta^{}_1 + |U^{}_{e 2}|^2 \right) \right.
\nonumber \\
\hspace{-0.2cm} & & \hspace{-0.2cm} + \zeta^{}_{21} \left[ \left(
|U^{}_{\tau 1}|^2 \Delta^{}_2 + |U^{}_{\tau 2}|^2 \Delta^{}_1 +
|U^{}_{e 3}|^2 \right) \cos\Phi^{}_{12} + {\cal J} \sin\Phi^{}_{12}
\right]
\nonumber \\
\hspace{-0.2cm} & & \hspace{-0.2cm} \left. + \zeta^{}_{31} \left[
\left( |U^{}_{\tau 1}|^2 \Delta^{}_3 + |U^{}_{\tau 3}|^2 \Delta^{}_1
+ |U^{}_{e 2}|^2 \right) \cos\Phi^{}_{13} - {\cal J}
\sin\Phi^{}_{13} \right]\right\} \; ,
\nonumber \\
16\pi^2 \frac{{\rm d}\Delta^{}_2}{{\rm d}t} \hspace{-0.2cm} & = &
\hspace{-0.2cm} +y^2_\tau \left\{ \xi^{}_{21} \left( |U^{}_{\tau
1}|^2 \Delta^{}_2 + |U^{}_{\tau 2}|^2 \Delta^{}_1 + |U^{}_{e 3}|^2
\right) - \xi^{}_{32} \left( |U^{}_{\tau 2}|^2 \Delta^{}_3 +
|U^{}_{\tau 3}|^2 \Delta^{}_2 + |U^{}_{e 1}|^2 \right) \right.
\nonumber \\
\hspace{-0.2cm} & & \hspace{-0.2cm} + \zeta^{}_{21} \left[ \left(
|U^{}_{\tau 1}|^2 \Delta^{}_2 + |U^{}_{\tau 2}|^2 \Delta^{}_1 +
|U^{}_{e 3}|^2 \right) \cos\Phi^{}_{12} + {\cal J} \sin\Phi^{}_{12}
\right]
\nonumber \\
\hspace{-0.2cm} & & \hspace{-0.2cm} \left. - \zeta^{}_{32} \left[
\left( |U^{}_{\tau 2}|^2 \Delta^{}_3 + |U^{}_{\tau 3}|^2 \Delta^{}_2
+ |U^{}_{e 1}|^2 \right) \cos\Phi^{}_{23} + {\cal J}
\sin\Phi^{}_{23} \right]\right\} \; ,
\nonumber \\
16\pi^2 \frac{{\rm d}\Delta^{}_3}{{\rm d}t} \hspace{-0.2cm} & = &
\hspace{-0.2cm} +y^2_\tau \left\{ \xi^{}_{31} \left( |U^{}_{\tau
1}|^2 \Delta^{}_3 + |U^{}_{\tau 3}|^2 \Delta^{}_1 + |U^{}_{e 2}|^2
\right) + \xi^{}_{32} \left( |U^{}_{\tau 2}|^2 \Delta^{}_3 +
|U^{}_{\tau 3}|^2 \Delta^{}_2 + |U^{}_{e 1}|^2 \right) \right.
\hspace{0.6cm}
\nonumber \\
\hspace{-0.2cm} & & \hspace{-0.2cm} + \zeta^{}_{31} \left[ \left(
|U^{}_{\tau 1}|^2 \Delta^{}_3 + |U^{}_{\tau 3}|^2 \Delta^{}_1 +
|U^{}_{e 2}|^2 \right) \cos\Phi^{}_{13} - {\cal J} \sin\Phi^{}_{13}
\right]
\nonumber \\
\hspace{-0.2cm} & & \hspace{-0.2cm} \left. + \zeta^{}_{32} \left[
\left( |U^{}_{\tau 2}|^2 \Delta^{}_3 + |U^{}_{\tau 3}|^2 \Delta^{}_2
+ |U^{}_{e 1}|^2 \right) \cos\Phi^{}_{23} + {\cal J}
\sin\Phi^{}_{23} \right]\right\} \; ,
\end{eqnarray}
where $\zeta^{}_{ij} \equiv 2m^{}_i m^{}_j/\Delta m^2_{ij}$,
$\cos\Phi^{}_{ij} \equiv {\rm Re}(U^{}_{\tau i} U^*_{\tau
j})^2/|U^{}_{\tau i} U^*_{\tau j}|^2$, $\sin\Phi^{}_{ij} \equiv {\rm
Im}(U^{}_{\tau i} U^*_{\tau j})^2/|U^{}_{\tau i} U^*_{\tau j}|^2$,
and the leptonic Jarlskog invariant ${\cal J}$ \cite{J} is defined
through
\begin{eqnarray}
{\rm Im}\left(U^{}_{\alpha i} U^{}_{\beta j} U^*_{\alpha j}
U^*_{\beta i}\right) = {\cal J} \sum_\gamma
\epsilon^{}_{\alpha\beta\gamma} \sum_k \epsilon^{}_{ijk} \;
\end{eqnarray}
with the Greek and Latin subscripts running over $(e, \mu, \tau)$
and $(1,2,3)$, respectively. Since the sign of $\zeta^{}_{ij}$ is
always the same as that of $\xi^{}_{ij}$, it is possible to adjust
the evolving direction of $\Delta^{}_3$ without much fine-tuning of
the other relevant parameters. Hence, similar to the Dirac neutrino case,
the RGE-triggered deviation of $\theta^{}_{23}$ from $45^\circ$
might be closely correlated with the neutrino mass ordering in
the Majorana case.

In Eq. (4) it should be noted that the two Majorana CP-violating
phases $\rho$ and $\sigma$ in the standard parametrization of $U$
affect the running behaviors of $\Delta^{}_i$ via $\cos\Phi^{}_{ij}$
and $\sin\Phi^{}_{ij}$. One should also note that ${\cal J} =
\left(\sin 2\theta^{}_{12} \sin 2\theta^{}_{13} \cos\theta^{}_{13}
\sin 2\theta^{}_{23} \sin\delta\right)/8$, which only depends on the
Dirac CP-violating phase $\delta$, measures the strength of CP
violation in neutrino oscillations. Therefore, $\delta \sim
270^\circ$ is especially favorable for significant CP-violating
effects in the lepton sector, no matter whether the massive
neutrinos are Dirac or Majorana particles.

\section{Numerical results for $\Delta^{}_i$, $\theta^{}_{23}$ and
$\delta$}

We proceed to numerically illustrate the effects of $\mu$-$\tau$
symmetry breaking regarding the PMNS matrix $U$ --- namely, the
quantities $\Delta^{}_i$ run from $\Delta^{}_i = 0$ (i.e.,
$\delta^{}_{23} = 45^\circ$ and $\delta = 270^\circ$) at
$\Lambda^{}_{\mu\tau} \sim 10^{14}$ GeV down to $\Lambda^{}_{\rm EW}
\sim 10^2$ GeV via the one-loop RGEs obtained in Eq. (3) or (4).
Given a proper value of $\tan\beta$, the values of $m^{}_1$, $\Delta
m^2_{21}$, $\Delta m^2_{31}$, $\theta^{}_{12}$ and $\theta^{}_{13}$
at $\Lambda^{}_{\mu\tau}$ should be carefully chosen such that the
best-fit results of $\Delta m^2_{21}$, $\Delta m^2_{31}$,
$\theta^{}_{12}$, $\theta^{}_{13}$, $\theta^{}_{23}$ and $\delta$ at
$\Lambda^{}_{\rm EW}$ as listed in Table 1 can all be reproduced to a good
degree of accuracy
\footnote{Note that the notations $\delta m^2 \equiv m^2_2 - m^2_1$
and $\Delta m^2 \equiv m^2_3 - (m^2_1 + m^2_2)/2$ have been used in
Ref. \cite{Fogli}. They are related with $\Delta m^2_{21}$ and
$\Delta m^2_{31}$ as follows: $\Delta m^2_{21} = \delta m^2$ and
$\Delta m^2_{31} = \Delta m^2 + \delta m^2/2$.}.
If this strategy is workable, then the deviation of $\theta^{}_{23}$
from $45^\circ$ and that of $\delta$ from $270^\circ$ will be purely
attributed to the RGE-triggered PMNS $\mu$-$\tau$ symmetry breaking
effects
\footnote{Note that the $\nu$fit group's best-fit results
\cite{NuFIT} are not taken into account in our numerical examples,
because they happen to correspond to the disfavored cases listed in
Table 1 (i.e., $\theta^{}_{23} < 45^\circ$ for the normal neutrino
mass ordering or $\theta^{}_{23} > 45^\circ$ for the inverted
ordering, in conflict with our expectations shown in Tables 2 and 3,
respectively).}.
\begin{table}[h]
\centering \caption{The best-fit values of $\Delta m^{2}_{21}$,
$\Delta m^{2}_{31}$, $\theta^{}_{12}$, $\theta^{}_{13}$,
$\theta^{}_{23}$ and $\delta$ obtained from two recent global
analyses of current neutrino oscillation data \cite{Fogli,Valle}.}
\vspace{5mm}
\begin{tabular}{l|l|cccccc} \hline
Reference & Mass ordering & $\Delta m^{2}_{21} ~({\rm eV}^2)$
& $\Delta m^{2}_{31} ~({\rm eV}^2)$
& $\theta^{}_{12}$ & $\theta^{}_{13}$ & $\theta^{}_{23}$ & $\delta$ \\
\hline
\multirow{2}{*}{Capozzi {\it et al} \cite{Fogli}} & Normal
& \multirow{2}{*}{$7.54 \times 10^{-5}$}
& $+2.47 \times 10^{-3}$ & \multirow{2}{*}{$33.7^\circ$}
& $8.8^\circ$ & $41.4^\circ$ & $250^\circ$ \\
& Inverted && $-2.34 \times 10^{-3}$
&& $8.9^\circ$ & $42.4^\circ$ & $236^\circ$ \\
\hline \multirow{2}{*}{Forero {\it et al} \cite{Valle}} & Normal &
\multirow{2}{*}{$7.60 \times 10^{-5}$} & $+2.48 \times 10^{-3}$ &
\multirow{2}{*}{$34.6^\circ$} & $8.8^\circ$ & $48.9^\circ$ &
$241^\circ$ \\ & Inverted && $-2.38 \times 10^{-3}$ && $8.9^\circ$ &
$49.2^\circ$ & $266^\circ$ \\ \hline
\end{tabular}
\end{table}

\subsection{Dirac neutrinos}

For simplicity, we fix $\tan\beta =31$ and input $m^{}_1 =0.1$ eV at
$\Lambda^{}_{\mu\tau}$, where $\Delta^{}_1$, $\Delta^{}_2$ and
$\Delta^{}_3$ are vanishing (or equivalently, $\theta^{}_{23} =
45^\circ$ and $\delta =90^\circ$), in our numerical calculations.
Table 2 shows the input and output values of all the relevant
parameters for two examples, which are based on the best-fit results
reported by Capozzi {\it et al} \cite{Fogli} and Forero {\it et al}
\cite{Valle}, respectively. Figure 1 illustrates how $\Delta^{}_i$
evolve in either example. Some comments and discussions are in
order.
\begin{table}[h]
\centering \caption{The RGE-triggered PMNS $\mu$-$\tau$ symmetry
breaking effects for Dirac neutrinos running from $\Delta^{}_i =0$
at $\Lambda^{}_{\mu\tau} \sim 10^{14}$ GeV down to $\Lambda^{}_{\rm
EW} \sim 10^2$ GeV in the MSSM with $\tan\beta = 31$.} \vspace{5mm}
\begin{tabular}{lllllll}
\hline \multirow{2}{*}{~ Parameter ~} && \multicolumn{2}{l}{Example
I (Capozzi {\it et al} \cite{Fogli})} && \multicolumn{2}{l}{Example
II (Forero {\it et al} \cite{Valle})} \\ \cline{3-4} \cline{6-7} &&
Input ($\Lambda_{\mu\tau}$) & Output ($\Lambda_{\rm EW}$) && Input
($\Lambda_{\mu\tau}$)
& Output ($\Lambda_{\rm EW}$) \\
\hline
~ $m^{}_{1} ~ ({\rm eV} )$ && 0.100 & 0.093 && 0.100 & 0.093 \\
~ $\Delta m^{2}_{21} ~ ({\rm eV}^2 )$
&& $1.82 \times 10^{-4}$ & $7.54 \times 10^{-5}$
&& $1.96 \times 10^{-4}$ & $7.60 \times 10^{-5}$ \\
~ $\Delta m^{2}_{31} ~ ({\rm eV}^2 )$
&& $-2.60 \times 10^{-3}$ & $-2.34 \times 10^{-3}$
&& $3.00 \times 10^{-3}$ & $2.48 \times 10^{-3}$ \\
~ $\theta^{}_{12}$ && $10.8^\circ$ & $33.6^\circ$ && $10.3^\circ$ &
$34.6^\circ$ \\ ~ $\theta^{}_{13}$ && $9.4^\circ$ & $8.9^\circ$ &&
$8.4^\circ$ &
$8.8^\circ$ \\
~ $\theta^{}_{23}$ && $45.0^\circ$ & $42.4^\circ$ && $45.0^\circ$ &
$48.4^\circ$ \\ ~ $\delta$ && $270^\circ$ & $236^\circ$ &&
$270^\circ$ & $237^\circ$
\\ \hline
~ ${\cal J}$ && $-0.015$ & $-0.029$ && $-0.013$ & $-0.029$ \\
~ $\Delta^{}_{1}$ && 0 & 0.053 && 0 & 0.114 \\
~ $\Delta^{}_{2}$ && 0 & $-0.141$ && 0 & 0.001 \\
~ $\Delta^{}_{3}$ && 0 & 0.088 && 0 & $-0.115$ \\
\hline
\end{tabular}
\end{table}

(1) Given the inverted neutrino mass ordering, the best-fit results
of the six neutrino oscillation parameters $\Delta m^2_{21}$,
$\Delta m^2_{31}$, $\theta^{}_{12}$, $\theta^{}_{13}$,
$\theta^{}_{23}$ and $\delta$ at $\Lambda^{}_{\rm EW}$ in Example I
\cite{Fogli} can successfully be reproduced from the proper inputs
at $\Lambda^{}_{\mu\tau}$. In this case $\theta^{}_{23}
(\Lambda^{}_{\rm EW})$ lies in the first octant, and $\theta^{}_{23}
(\Lambda^{}_{\mu\tau}) - \theta^{}_{23} (\Lambda^{}_{\rm EW}) \simeq
2.6^\circ$ holds thanks to the RGE running effect. At the same time, we
obtain $\delta (\Lambda^{}_{\mu\tau}) - \delta (\Lambda^{}_{\rm EW})
\simeq 34^\circ$. Hence the RGE evolution can also provide a
resolution to the quadrant of $\delta$.

(2) In contrast, only the normal neutrino mass ordering allows us to
obtain $\theta^{}_{23}(\Lambda^{}_{\rm EW}) \simeq 48.4^\circ$ from
$\theta^{}_{23}(\Lambda^{}_{\mu\tau}) = 45^\circ$ via the RGE
evolution as shown in Example II \cite{Valle}. Moreover, we obtain
$\delta(\Lambda^{}_{\rm EW}) \simeq 237^\circ$ from
$\delta(\Lambda^{}_{\mu\tau}) = 270^\circ$, and this result is also
consistent very well with the corresponding best-fit value $\delta
\simeq 241^\circ$ as listed in Table 1. The future experimental data
will only verify one of the above two possibilities for the octant
of $\theta^{}_{23}$, but it will be interesting to test the expected
correlation between the neutrino mass ordering and the deviation of
$\theta^{}_{23}$ (or $\delta$) from $45^\circ$ (or $270^\circ$).

(3) Figure 1 shows the behaviors of three PMNS $\mu$-$\tau$
asymmetries $\Delta^{}_i$ evolving from $\Lambda^{}_{\mu\tau}$ down
to $\Lambda^{}_{\rm EW}$ for the two examples under discussion. In
view of $\Delta^{}_3 = \cos^2\theta^{}_{13} \cos 2\theta^{}_{23}$ in
Eq. (2), one must have $\Delta^{}_3 (\Lambda^{}_{\rm EW}) > 0$ for
$\theta^{}_{23} (\Lambda^{}_{\rm EW}) < 45^\circ$ in Example I, and
$\Delta^{}_3 (\Lambda^{}_{\rm EW}) < 0$ for $\theta^{}_{23}
(\Lambda^{}_{\rm EW}) > 45^\circ$ in Example II. In comparison, the
running behaviors of $\Delta^{}_1$ and $\Delta^{}_2$ are not so
straightforward, because they depend on all the three flavor mixing
angles and the CP-violating phase $\delta$. But $\Delta^{}_1 +
\Delta^{}_2 + \Delta^{}_3 =0$ holds at any energy scale between
$\Lambda^{}_{\rm EW}$ and $\Lambda^{}_{\mu\tau}$, as one can see in
Figure 1.

\subsection{Majorana neutrinos}

In this case we simply fix $\tan\beta =30$ and input $m^{}_1 =0.1$
eV at $\Lambda^{}_{\mu\tau}$, where $\Delta^{}_1 = \Delta^{}_2 =
\Delta^{}_3 =0$ holds, in our numerical calculations. Table 3 is a
brief summary of the input and output values of all the relevant
parameters for Example I \cite{Fogli} and Example II \cite{Valle},
respectively. In addition, Figure 2 illustrates how the three PMNS
$\mu$-$\tau$ asymmetries evolve from $\Lambda^{}_{\mu\tau}$ down to
$\Lambda^{}_{\rm EW}$ in either example.

Although the present case involves two extra CP-violating phases
$\rho$ and $\sigma$, the running behaviors of $\Delta^{}_i$ in
Figure 2 are quite similar to those in Figure 1. Of course, one has
to adjust the initial values of $\rho$ and $\sigma$ at
$\Lambda^{}_{\mu\tau}$ in a careful way, such that the best-fit
results of the six neutrino oscillation parameters can correctly be
reproduced at $\Lambda^{}_{\rm EW}$. We find that it is really
possible to resolve the octant of $\theta^{}_{23}$ and the quadrant
of $\delta$ at the same time via radiative PMNS $\mu$-$\tau$
symmetry breaking. Very similar to the Dirac neutrino case, the
RGE-triggered deviation of $\theta^{}_{23}$ from $45^\circ$ in the
Majorana case is also closely correlated with the neutrino mass
ordering. Namely, $\theta^{}_{23} \simeq 42.4^\circ$ reported in
Ref. \cite{Fogli} (or $\theta^{}_{23} \simeq 48.9^\circ$ reported in
Ref. \cite{Valle}) at $\Lambda^{}_{\rm EW}$ can evolve from
$\theta^{}_{23} = 45^\circ$ at $\Lambda^{}_{\mu\tau}$ only when the
neutrino masses have an inverted (or normal) ordering.
\begin{table}[h]
\centering \caption{The RGE-triggered PMNS $\mu$-$\tau$ symmetry
breaking effects for Majorana neutrinos running from $\Delta^{}_i
=0$ at $\Lambda^{}_{\mu\tau} \sim 10^{14}$ GeV down to
$\Lambda^{}_{\rm EW} \sim 10^2$ GeV in the MSSM with $\tan\beta =
30$.} \vspace{5mm}
\begin{tabular}{lllllll}
\hline \multirow{2}{*}{~ Parameter ~} && \multicolumn{2}{l} {Example
I (Capozzi {\it et al} \cite{Fogli})} && \multicolumn{2}{l}{Example
II (Forero {\it et al} \cite{Valle})} \\ \cline{3-4} \cline{6-7} &&
Input ($\Lambda_{\mu\tau}$) & Output ($\Lambda_{\rm EW}$)
&& Input ($\Lambda_{\mu\tau}$) & Output ($\Lambda_{\rm EW}$) \\
\hline
~ $m^{}_{1} ~ ({\rm eV} )$ && 0.100 & 0.087 && 0.100 & 0.087 \\
~ $\Delta m^{2}_{21} ~ ({\rm eV}^2 )$ && $1.70 \times 10^{-4}$
& $7.54 \times 10^{-5}$ && $2.12 \times 10^{-4}$
& $7.60 \times 10^{-5}$ \\
~ $\Delta m^{2}_{31} ~ ({\rm eV}^2 )$ && $-2.98 \times 10^{-3}$
& $-2.34 \times 10^{-3}$ && $3.50 \times 10^{-3}$
& $2.48 \times 10^{-3}$ \\
~ $\theta^{}_{12}$ && $35.2^\circ$ & $33.7^\circ$ && $32.1^\circ$ &
$34.6^\circ$ \\
~ $\theta^{}_{13}$ && $10.1^\circ$ & $8.9^\circ$ && $6.9^\circ$ &
$8.8^\circ$ \\
~ $\theta^{}_{23}$ && $45.0^\circ$ & $42.4^\circ$ && $45.0^\circ$ &
$48.9^\circ$ \\
~ $\delta$ && $270^\circ$ & $236^\circ$ && $270^\circ$ & $241^\circ$
\\ ~ $\rho$ && $-82^\circ$ & $-66^\circ$ && $-76^\circ$ &
$-45^\circ$
\\
~ $\sigma$ && $19^\circ$ & $27^\circ$ && $17^\circ$ & $29^\circ$ \\
\hline
~ ${\cal J}$ && $-0.040$ & $-0.029$ && $-0.027$ & $-0.030$ \\
~ $\Delta^{}_{1}$ && 0 & 0.054 && 0 & 0.111 \\
~ $\Delta^{}_{2}$ && 0 & $-0.142$ && 0 & 0.022 \\
~ $\Delta^{}_{3}$ && 0 & 0.088 && 0 & $-0.133$ \\
\hline
\end{tabular}
\end{table}

In view of the fact that the present best-fit results of
$\theta^{}_{23}$ and $\delta$ are still quite preliminary, we
foresee that they must undergo some changes before they are well
determined by the more precise experimental data in the near future.
Hence our numerical analysis is not targeted for a complete
parameter-space exploration but mainly for the purpose of
illustration \cite{Ohlsson}. Its outcome supports our original
conjecture: the slight $\mu$-$\tau$ symmetry breaking behind the
observed pattern of lepton flavor mixing can originate from the RGE
evolution from a superhigh flavor symmetry scale down to the
electroweak scale. Note that there are two adjustable unknown
parameters in our calculations: the absolute neutrino mass $m^{}_1$
and the MSSM parameter $\tan\beta$. Once $m^{}_1$ is experimentally
determined and $\tan\beta$ is theoretically fixed, for example, it
will be interesting to see whether one can still resolve the octant
of $\theta^{}_{23}$ and the quadrant of $\delta$ with the help of
radiative PMNS $\mu$-$\tau$ symmetry breaking effects.

We admit that the present best-fit result $\delta \sim 270^\circ$
remains too preliminary. In fact, there is not any nontrivial region
associated with the allowed values of $\delta$ at the $2\sigma$
level \cite{Fogli,Valle}. Hence it also makes sense to look at the
RGE-triggered corrections to $\theta^{}_{23} = 45^\circ$ and $\delta
= 90^\circ$ for the energy scale to evolve from
$\Lambda^{}_{\mu\tau}$ down to $\Lambda^{}_{\rm EW}$. This
possibility has already been discussed in some literature (see,
e.g., Refs. \cite{Ohlsson,Mei}). Once the CP-violating phase
$\delta$ is measured or constrained to a better degree of accuracy
in the near future, it will be possible to examine whether the
quantum corrections can really accommodate the observed effect of
PMNS $\mu$-$\tau$ symmetry breaking or not.

\section{Summary and further discussions}

To summarize, we have conjectured that the PMNS $\mu$-$\tau$
permutation symmetry is exact at a superhigh energy scale
$\Lambda^{}_{\mu\tau} \sim 10^{14}$ GeV, where the origin of
neutrino masses and flavor mixing has a good dynamic reason, and its
slight breaking happens via the RGE running down to the electroweak
scale $\Lambda^{}_{\rm EW} \sim 10^2$ GeV. This idea is particularly
interesting in the sense that it can help resolve the octant of
$\theta^{}_{23}$ and the quadrant of $\delta$ at the same time
thanks to radiative PMNS $\mu$-$\tau$ symmetry breaking in the MSSM.
In fitting current neutrino oscillation data we have found that the
RGE-triggered deviation of $\theta^{}_{23}$ from $45^\circ$ is
uniquely correlated with the neutrino mass ordering: $\theta^{}_{23}
\simeq 42.4^\circ$ \cite{Fogli} (or $\theta^{}_{23} \simeq
48.9^\circ$ \cite{Valle}) at $\Lambda^{}_{\rm EW}$ can naturally
originate from $\theta^{}_{23} = 45^\circ$ at $\Lambda^{}_{\mu\tau}$
if the neutrino mass ordering is inverted (or normal). Accordingly,
the preliminary best-fit results of $\delta$ at $\Lambda^{}_{\rm
EW}$ can also evolve from $\delta = 270^\circ$ at
$\Lambda^{}_{\mu\tau}$. Such remarkable findings are independent of
any specific models of neutrino mass generation and lepton flavor
mixing, and they will soon be tested in the upcoming neutrino
oscillation experiments.

Note that some previous studies of the RGE evolution of lepton
flavor mixing parameters have more or less involved the $\mu$-$\tau$
symmetry breaking effects \cite{Zhou}. In this connection a few
constant neutrino mixing patterns which possess $|U^{}_{\mu i}| =
|U^{}_{\tau i}|$, such as the bimaximal \cite{BM} and tri-bimaximal
\cite{TBM} ones with $\theta^{}_{13} =0^\circ$ and $\theta^{}_{23}
=45^\circ$, have been assumed at a superhigh energy scale; and their
RGE running behaviors have been investigated mainly to see whether a
finite $\theta^{}_{13}$ can be radiatively generated at low energies
\cite{Mei}. The closest example of this kind should be the work
\cite{Zhang} on radiative corrections to the tetra-maximal neutrino
mixing pattern \cite{TM}, in which $\theta^{}_{13} \simeq
8.4^\circ$, $\theta^{}_{23} = 45^\circ$ and $\delta = 90^\circ$ or
$270^\circ$ have been predicted. Our present work is different from
the previous ones in several aspects: (a) it is not subject to any
explicit neutrino mixing pattern; (b) it focuses on the PMNS
$\mu$-$\tau$ asymmetries $\Delta^{}_i$ and its RGE evolution; (c) it
provides a reasonable resolution to the octant of $\theta^{}_{23}$
by attributing it to the PMNS $\mu$-$\tau$ symmetry breaking effect;
(d) it may also resolve the quadrant of $\delta$ in a similar way.
All in all, we have established the RGE connection between a given
neutrino mass model with the exact $\mu$-$\tau$ symmetry at
superhigh energies and the neutrino oscillation parameters at low
energies. Such a connection is expected to be very useful for
neutrino phenomenology in the era of precision measurements.

\vspace{0.5cm}

This work was supported in part by the National Natural
Science Foundation of China under grant No. 11105113 (S.L.) and
grant No. 11135009 (Z.Z.X.).

\newpage

\begin{figure}
\begin{center}
\vspace{0cm}
\includegraphics[scale=0.66, angle=0, clip=0]{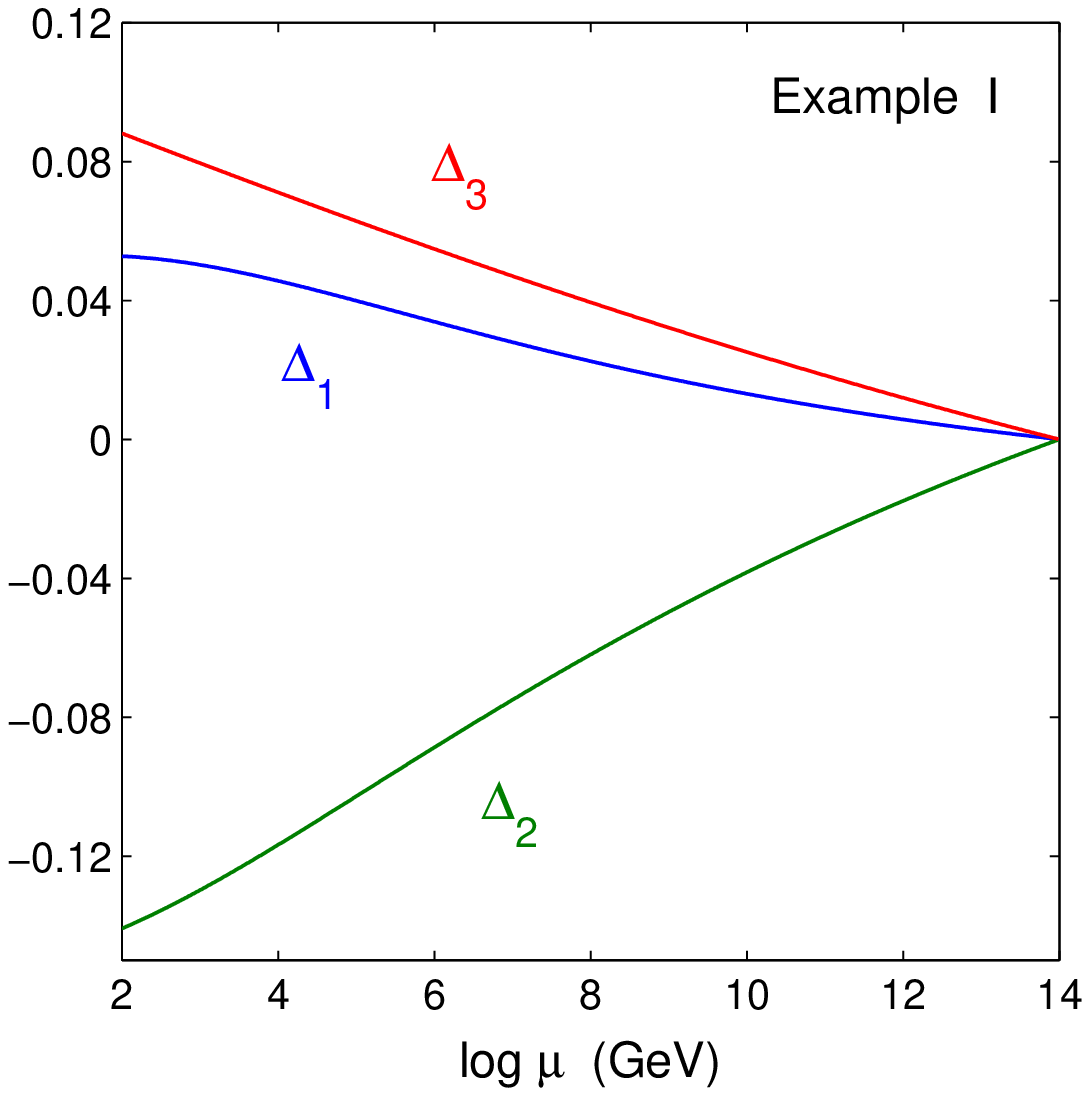}
\includegraphics[scale=0.66, angle=0, clip=0]{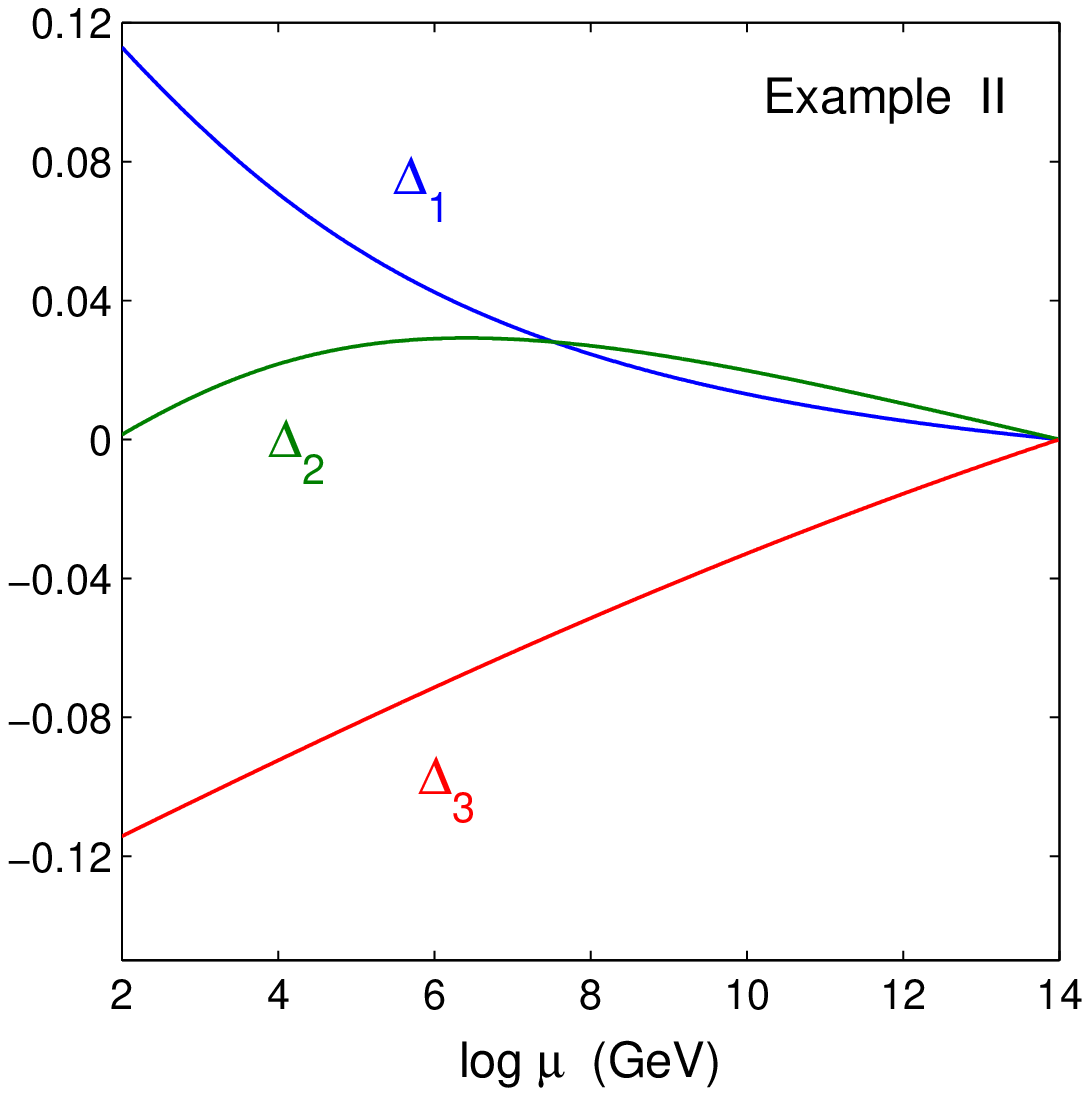}
\vspace{-0.6cm} \caption{The RGE-triggered $\mu$-$\tau$ symmetry
breaking effects for Dirac neutrinos running from $\Delta^{}_i =0$
at $\Lambda^{}_{\mu\tau} \sim 10^{14}$ GeV down to $\Lambda^{}_{\rm
EW} \sim 10^2$ GeV in the MSSM with $\tan\beta = 31$.}
\end{center}
\end{figure}

\begin{figure}
\begin{center}
\vspace{0cm}
\includegraphics[scale=0.66, angle=0, clip=0]{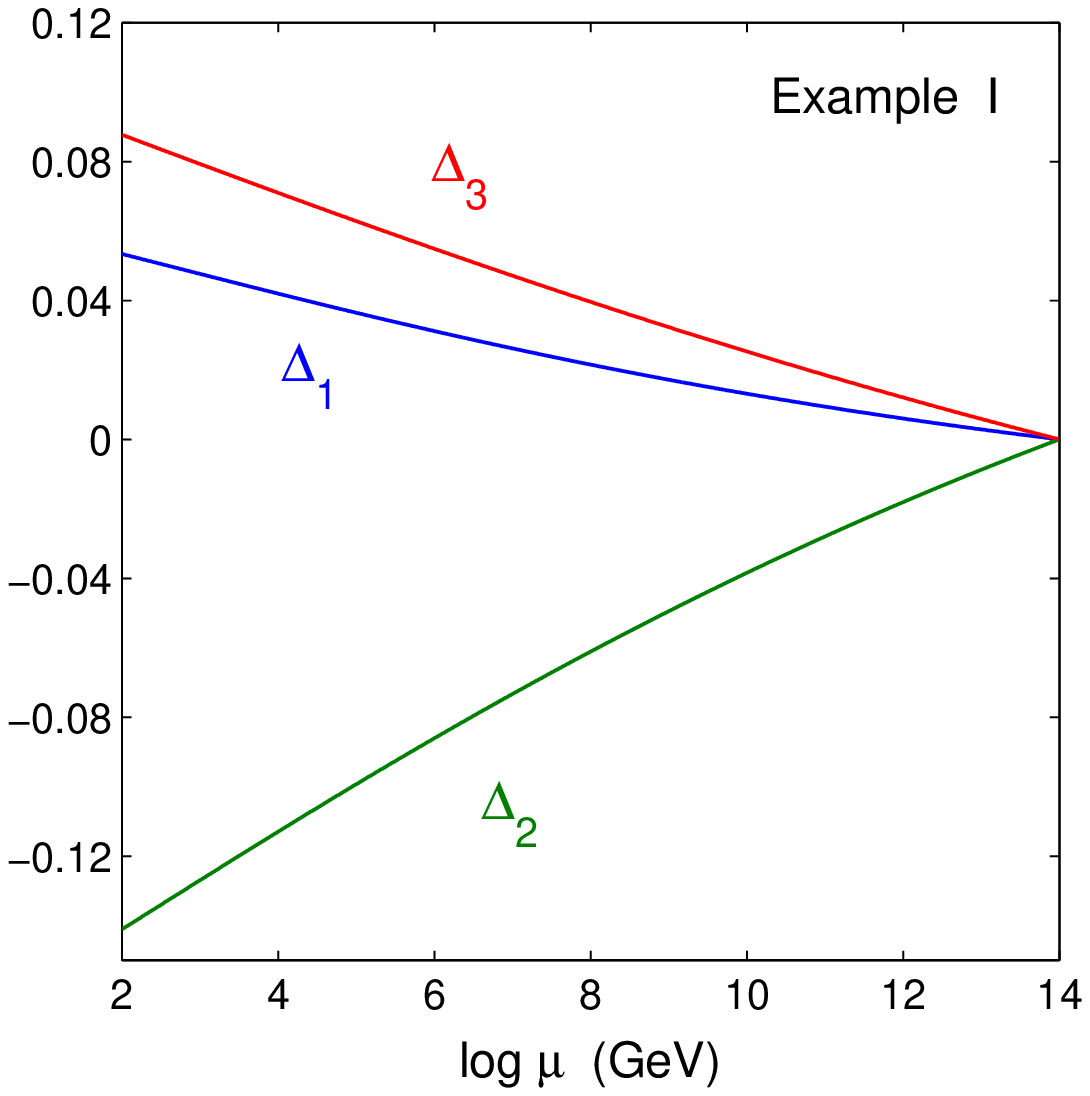}
\includegraphics[scale=0.66, angle=0, clip=0]{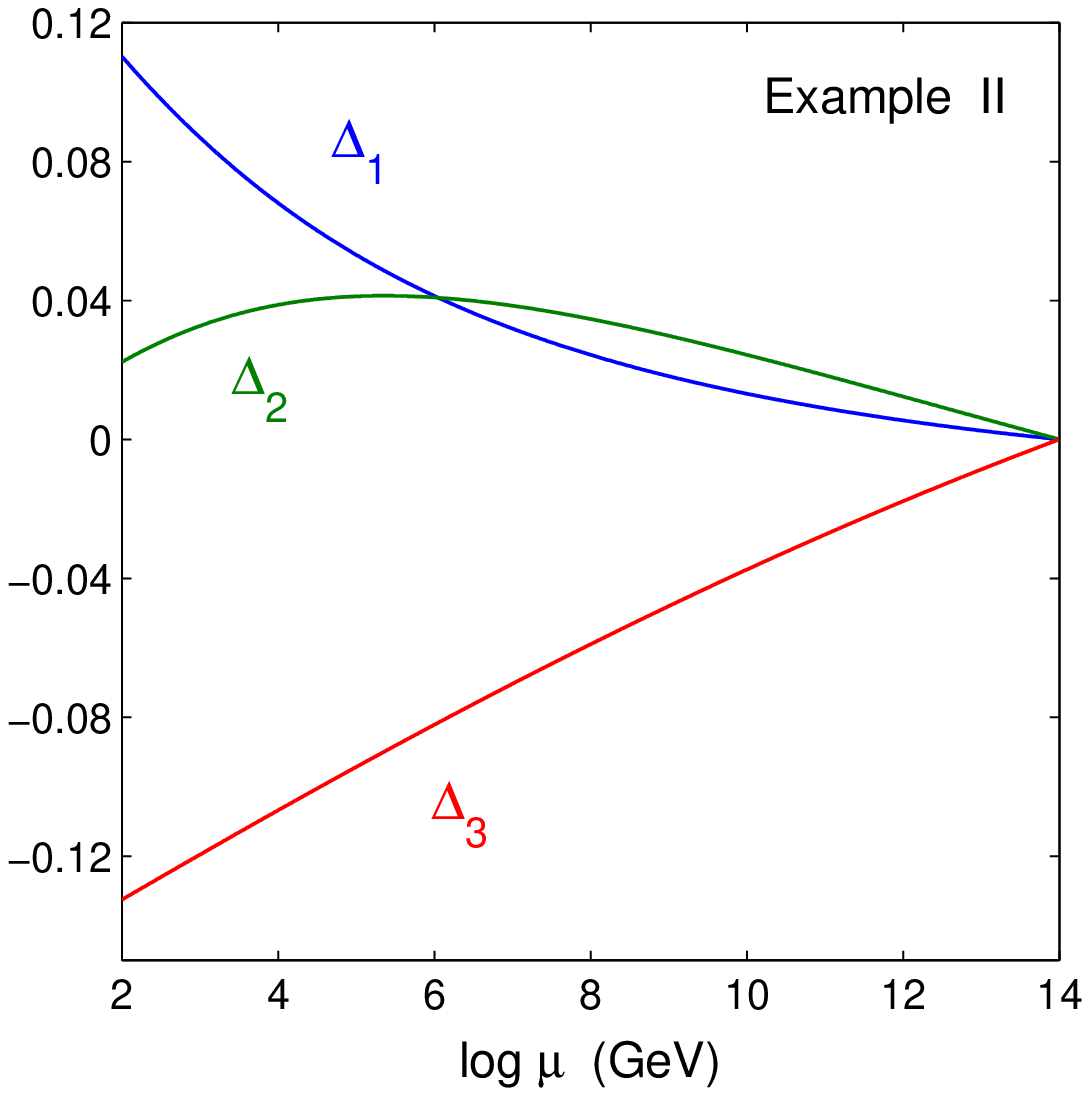}
\vspace{-0.6cm}
\caption{The RGE-triggered $\mu$-$\tau$ symmetry breaking effects for
Majorana neutrinos running from $\Delta^{}_i =0$ at
$\Lambda^{}_{\mu\tau} \sim 10^{14}$ GeV
down to $\Lambda^{}_{\rm EW} \sim 10^2$ GeV in the MSSM with
$\tan\beta = 30$.}
\end{center}
\end{figure}


\newpage

\begin{thebibliography}{99}
\bibitem{SK} Y. Fukuda {\it et al.} (Super-Kamiokande Collaboration),
Phys. Rev. Lett. {\bf 81}, 1562 (1998).

\bibitem{DYB} F.P. An {\it et al.} (Daya Bay Collaboration),
Phys. Rev. Lett. {\bf 108}, 171803 (2012).

\bibitem{Fogli} F. Capozzi, G.L. Fogli, E. Lisi, A. Marrone, D.
Montanino, and A. Palazzo, Phys. Rev. D {\bf 89}, 093018 (2014).

\bibitem{Valle} D.V. Forero, M. Tortola, and J.W.F. Valle,
arXiv:1405.7540.

\bibitem{T2K} K. Abe {\it et al.} (T2K Collaboration), Phys. Rev. Lett.
{\bf 112}, 061802 (2014).

\bibitem{MNS} Z. Maki, M. Nakagawa, and S. Sakata, Prog. Theor.
Phys. {\bf 28}, 870 (1962); B. Pontecorvo, Sov. Phys. JETP {\bf 26},
984 (1968).

\bibitem{XZ} Z.Z. Xing and S. Zhou, Phys. Lett. B {\bf 737}, 196
(2014).

\bibitem{CKM} N. Cabibbo, Phys. Rev. Lett. {\bf 10}, 531 (1963);
M. Kobayashi and T. Maskawa, Prog. Theor. Phys. {\bf 49}, 652 (1973).

\bibitem{SS} P. Minkowski, Phys. Lett. B {\bf 67}, 421 (1977);
T. Yanagida, in {\it Proceedings of the Workshop on Unified Theory
and the Baryon Number of the Universe}, edited by O. Sawada and A.
Sugamoto (KEK, Tsukuba, 1979), p. 95; M. Gell-Mann, P. Ramond, and
R. Slansky, in {\it Supergravity}, edited by P. van Nieuwenhuizen
and D. Freedman (North Holland, Amsterdam, 1979), p. 315; S.L.
Glashow, in {\it Quarks and Leptons}, edited by M. L$\acute{\rm
e}$vy {\it et al.} (Plenum, New York, 1980), p. 707; R.N. Mohapatra
and G. Senjanovic, Phys. Rev. Lett. {\bf 44}, 912 (1980).

\bibitem{Flavor} For a flavor symmetry model which simultaneously
predicts $\theta^{}_{23} = 45^\circ$ and $\delta = 270^\circ$, see,
e.g., R.N. Mohapatra and C.C. Nishi, Phys. Rev. D {\bf 86}, 073007
(2012); P.M. Ferreira, W. Grimus, L. Lavoura, and P.O. Ludl, JHEP
{\bf 1209}, 128 (2012);  H.J. He and X.J. Xu, Phys. Rev. D {\bf 86} 111301 (2012); 
F. Feruglio, C. Hagedorn, and R. Ziegler,
JHEP {\bf 1307}, 027 (2013); Eur. Phys. J. C {\bf 74}, 2753 (2014);
W. Grimus and L. Lavoura, Fortsch. Phys. {\bf 61}, 535 (2013); G.J.
Ding, S.F. King, C. Luhn, and A.J. Stuart, JHEP {\bf 1305}, 084
(2013); G.J. Ding, S.F. King, and A.J. Stuart, JHEP {\bf 1312}, 006
(2013); C.C. Li and G.J. Ding, Nucl. Phys. B {\bf 881}, 206 (2014);
G.J. Ding and Y.L. Zhou, arXiv:1312.5222; JHEP {\bf 1406}, 023
(2014); G.J. Ding and S.F. King, Phys. Rev. D {\bf 89}, 093020
(2014).

\bibitem{XZZ} See, e.g., Z.Z. Xing, H. Zhang, and S. Zhou,
Phys. Rev. D {\bf 86}, 013013 (2012); J. Elias-Miro, J.R. Espinosa,
G.F. Giudice, G. Isidori, A. Riotto, and A. Strumia, Phys. Lett.
B {\bf 709}, 222 (2012).

\bibitem{PDG} K.A. Olive {\it et al.} (Particle Data Group),
Chin. Phys. C {\bf 38}, 090001 (2014).

\bibitem{XZ2008} Z.Z. Xing and S. Zhou, Phys. Lett. B {\bf 666},
166 (2008).

\bibitem{Luo} S. Luo, Phys. Rev. D {\bf 85}, 013006 (2012).

\bibitem{J} C. Jarlskog, Phys. Rev. Lett. {\bf 55}, 1039 (1985);
Z. Phys. C {\bf 29}, 491 (1985).

\bibitem{NuFIT} M.C. Gonzalez-Garcia, M. Maltoni, J. Salvado, and
T. Schwetz, JHEP {\bf 12}, 123 (2012).

\bibitem{Ohlsson} A more detailed analysis of the uncertainties of
CP-violating phases and flavor mixing angles in their RGE evolution
can be found in: T. Ohlsson, H. Zhang, and S. Zhou, Phys. Rev. D {\bf 87},
013012 (2013).

\bibitem{Mei} See, e.g., S. Antusch, J. Kersten, M. Lindner, and M.
Ratz, Phys. Lett. B {\bf 544}, 1 (2002); J.W. Mei and Z.Z. Xing,
Phys. Rev. D {\bf 70}, 053002 (2004); S. Luo and Z.Z. Xing, Phys.
Lett. B {\bf 632}, 341 (2006); S. Goswami, S.T. Petcov, S. Ray, and
W. Rodejohann, Phys. Rev. D {\bf 80}, 053013 (2009).

\bibitem{Zhou} For a recent review with extensive references,
see: T. Ohlsson and S. Zhou, arXiv:1311.3846.

\bibitem{BM} See, e.g., F. Vissani, arXiv:hep-ph/9708483;
V. Barger, S. Pakvasa, T.J. Weiler, and K. Whisnant, Phys. Lett. B
{\bf 437}, 107 (1998); H. Fritzsch and Z.Z. Xing,
Phys. Lett. B {\bf 440}, 313 (1998).

\bibitem{TBM} P.F. Harrison, D.H. Perkins, and W.G. Scott, Phys. Lett.
B {\bf 530}, 167 (2002); Z.Z. Xing, Phys. Lett. B {\bf 533}, 85
(2002); P.F. Harrison and W.G. Scott, Phys. Lett. B {\bf 535}, 163
(2002); X.G. He and A. Zee, Phys. Lett. B {\bf 560}, 87 (2003).

\bibitem{Zhang} H. Zhang and S. Zhou, Phys. Lett. B {\bf 704}, 296
(2011).

\bibitem{TM} Z.Z. Xing, Phys. Rev. D {\bf 78}, 011301 (2008).

\end{thebibliography}
\end{document}